# Deep Learning-based Feature Discovery for Decoding Phenotypic Plasticity in Pediatric High-Grade Gliomas Single-Cell Transcriptomics

Abicumaran Uthamacumaran[1,2,3]

[1]Department of Experimental Surgery, McGill University, Montreal, Canada
[2]Department of Physics (Alumni), Concordia University, Montreal, Canada
[2]Department of Psychology (Alumni), Concordia University, Montreal, Canada
[3]Oxford Immune Algorithmics, Reading, UK

**ABSTRACT**

Advancements in AI-powered systems medicine have revolutionized biomarker discovery through emergent and explainable features. By use of complex network dynamics and graph-based machine learning, we identified critical determinants of lineage-specific plasticity across the single-cell transcriptomics of pediatric high-grade glioma (pHGGs) subtypes: IDHWT glioblastoma and K27M-mutant glioma. Our study identified network interactions regulating glioma morphogenesis via the tumor-immune microenvironment, including neurodevelopmental programs, calcium dynamics, iron metabolism, metabolic reprogramming, and feedback loops between MAPK/ERK and WNT signaling. These relationships highlight the emergence of a hybrid spectrum of cellular states navigating a disrupted neuro-differentiation hierarchy. We identified transition genes such as DKK3, NOTCH2, GATAD1, GFAP, and SEZ6L in IDHWT glioblastoma, and H3F3A, ANXA6, HES6/7, SIRT2, FXYD6, PTPRZ1, MEIS1, CXXC5, and NDUFAB1 in K27M subtypes. We also identified MTRNR2L1, GAPDH, IGF2, FKBP variants, and FXYD7 as transition genes that influence cell fate decision-making across both subsystems. Our findings suggest pHGGs are developmentally trapped in states exhibiting maladaptive behaviors, and hybrid cellular identities. In effect, tumor heterogeneity (metastability) and plasticity emerge as stress-response patterns to immune-inflammatory microenvironments and oxidative stress. Furthermore, we show that pHGGs are steered by developmental trajectories from radial glia predominantly favoring neocortical cell fates, in telencephalon and prefrontal cortex (PFC) differentiation. By addressing underlying patterning processes and plasticity networks as therapeutic vulnerabilities, our findings provide precision medicine strategies aimed at modulating glioma cell fates and overcoming therapeutic resistance. We suggest transition therapy toward neuronal-like lineage differentiation as a potential therapy to help stabilize pHGG plasticity and aggressivity.

**INTRODUCTION**

Pediatric high-grade gliomas (pHGGs), including K27M-mutant diffuse intrinsic pontine gliomas (DIPG) and IDH wild-type (IDHWT) glioblastomas, are lethal childhood brain tumors, with dismal survival rates and no effective diagnostics, prevention, or treatments (Roberts et al., 2023; An al., 2024). These malignancies arise from disrupted neurodevelopmental programs, rooted in neural stem cells (NSCs) (i.e., radial glia within the subventricular zone (SVZ)) or their progenitor cells (Ocasio et al., 2023; Wang et al., 2024). These cells of origin harbor driver mutations that interfere with differentiation hierarchies, leading to a stem-like state characterized by emergent, maladaptive behaviors such as therapy resistance, immune evasion, neuromodulator hijacking (i.e., neuron-glioma interactions), metabolic reprogramming, and aggressive invasion (Jessa et al., 2019; Couturier et al., 2020; Winkler et al., 2023; Baig and Winkler, 2024; Wang et al., 2024). The crux of these maladaptive traits lies in epigenetic or phenotypic plasticity (Davies et al., 2023; Mehta and Stanger, 2024). In addition, tumor invasion and the blood-brain barrier complicate treatment delivery, limiting therapeutic efficacy. This "blocked differentiation", or "developmental arrest" exacerbated by epigenetic and genetic aberrations, promotes tumor progression, making current treatment strategies largely palliative (Jessa et al., 2019). The poor patient outcomes and high morbidity highlight the urgent need for innovative approaches and targeted therapies. However, the complexity of pHGGs is compounded by their tumor microenvironment (TME) evolution and metastable behaviors, akin to *neurodevelopmental identity disorders* at the cellular scale. Phenotypic plasticity serves as an evolvability engine steering the collective tumor ecosystem to navigate a fluid, hybrid spectrum of cellular identities (Davies et al., 2023; Mehta and Stanger, 2024). As such, deep learning algorithms and systems medicine approaches can help recognize patterns in multi-omics data, such as single-cell transcriptomics, to reveal the underlying complex networks driving the plasticity dynamics (Senft et al., 2019; Groves and Quaranta, 2023). *Systems medicine* investigates complex dynamics and (emergent) collective behaviors in tumor ecosystems through transdisciplinary approaches such as dynamical systems theory, network medicine, and data science (Du et al., 2015; Wang et al., 2021[a]; Groves and Quaranta, 2023). By emergent behaviors, we refer to the collective agency (intelligence) that arises from the nonlinear, many-part interactions of the system steered towards a common teleonomy (goal). Examples of emergent behaviors include the self-organization of beehives, flocking of birds, swarm intelligence, oscillations of neural networks, and the collective dynamics of tumor ecosystems (Ladyman and Wiesner, 2020).

We hypothesized that by comparing K27M DIPG and IDHWT glioblastoma, graph-based deep learning approaches combined with network medicine can uncover shared and distinct pathways modulating pHGG cell fate choices, thus, guiding precision therapies aimed at cell fate reprogramming these cells toward differentiated, less malignant states (i.e., cancer reversal). Deep learning predicts nonlinear patterns and complex, emergent relationships unanticipated by traditional statistical methods. Its integration with single-cell transcriptomics enables the inference of causal patterns and predictive biomarkers, which may identify therapeutic vulnerabilities in pHGGs. Generative adversarial networks (GANs) and variational autoencoders (VAEs), alongside graph-based neural networks (GNNs) like graph attention networks (GATs) and graph convolutional networks (GCNs), are particularly suited to analyzing emergent interactions in complex, high-dimensional data which can be represented as graph-theoretic networks (Marouf et al., 2020; Xu et al., 2020; Grønbech et al., 2020; Wang et al., 2021[a,b,c]; Chen & Liu, 2022; Abadi et al., 2023; Choi et al., 2023; Yuan et al., 2024; Yang et al., 2024). We predicted that these Deep learning models can infer the long-term (asymptotic) behavioral patterns of cell fate decision-making (i.e., attractors). Attractors refer to the emergent patterns in gene expression state space (epigenetic landscape) to which the dynamics of the system, or cell fate trajectories, are bound (Li and Wang, 2013). By inferring attractor dynamics, the algorithms can detect *critical transitions* (i.e., the tipping points at which the state transitions from one phenotype or linage to another) (Wang and Chen, 2020; Shin and Cho, 2023; Barcenas et al., 2024). Using Hopfield networks as a baseline for attractor inference, we aim to compare how these Deep learning algorithms capture metastable behaviors and cell fate decisions (Fard et al., 2016; Guo and Zheng, 2017). Moreover, these deep-learning approaches are vastly underexplored in pediatric tumors.

**METHODS**

**Gene Expression Datasets:** Gene expression matrices were obtained through high-throughput single-cell RNA sequencing (scRNA-seq), resulting in count matrices represented in TPM (transcripts per million). Quality control retained cells with at least 200 detected genes while excluding genes present in fewer than three cells. The datasets included Neftel et al. (1,943 cells from eight pediatric IDH-WT GBM patients) and Filbin et al. (2,458 cells from six H3K27M DIPG patients) (Filbin et al., 2018; Neftel et al., 2019). Log-normalized counts matrices were processed further by the artificial neural networks (ANNs).

**Hopfield Neural Network:** The Hopfield network is an associative memory model built using the numpy library, adapted from (Guo and Zheng, 2017). It was trained using the Hebbian learning rule, where weights (W) are calculated as the outer product of each binary pattern (p) and summed across all patterns: W=∑p⊗p. Diagonal weights were set to zero to prevent self-connections. The network's energy for a state vector (S) was computed using $E = -\frac{1}{2} S^T \cdot W \cdot S$. These state update followed the asynchronous rule: $S_{t+1} = sign(W \cdot S_t)$. Normalized gene expression data was binarized (1 or −1) as input patterns. Clustering was performed using the K-Means algorithm from sklearn, with the optimal number of clusters (K) determined via silhouette scores for k∈[2,10]. Random states (1000) were generated to map energy landscapes, and Differentially Expressed Genes (DEGs) were identified using scipy's t-test or Mann-Whitney U test depending on data normality, to compare clusters and energy ranges for the cluster differentiation. The network had one layer with weights representing pairwise interactions, no explicit hidden layers, and learning was unsupervised, parameter-free beyond normalization and binarization. Results revealed gene clusters along a two-dimensional energy landscapes. The energy in a Hopfield network is unitless, as it is a dimensionless scalar representing the network's stability.

**Variational Autoencoder (VAE):** Implemented in torch (PyTorch). A VAE with 128 hidden neurons and 32 latent dimensions was trained over 100 epochs using Adam optimizer (learning rate = 0.001), a binary cross-entropy loss, and Kullback-Leibler (KL)-divergence. The latent space representation was extracted from the VAE, and clusters were identified using KMeans clustering (k=3). Transition genes were identified using the Mann-Whitney U test to compare cluster-specific gene expression profiles against other clusters. DEGs with a p-value < 0.05 were further screened. Feature importance was analyzed using a Random Forest Classifier from scikit-learn (n=100 decision trees). Random forests were trained to distinguish each cluster, and the top 30 most important genes (explainable features) were extracted. The VAE used a combined loss function comprising binary cross-entropy for reconstruction and KL-divergence for regularization.

**Generative Adversarial Network (GAN):** The scGAN framework consists of the VAE as a Generator and a Discriminator. The generator learns to encode input data into a latent space and reconstruct it while capturing its variability. The discriminator distinguishes between true batch labels and generated outputs, ensuring robust latent representations. The VAE was implemented with PyTorch, consisting of a 64-unit hidden layer, 10 latent dimensions, and an input feature dimension equal to the number of genes. The model incorporated batch information as an additional input and used reparameterization to map data into latent space. The GANTrainer class orchestrated training over 30 epochs using the Adam optimizer with a learning rate of 0.001 and epsilon of 1×10 E-8. Reconstruction loss (mean squared error; MSE) and KL divergence were optimized for the VAE, while the discriminator used cross-entropy loss to distinguish batches. Latent representations were clustered using KMeans (10 clusters, random state 42), followed by visualization with nonlinear dimensionality reduction: t-SNE or UMAP via the Scanpy library. Clustering robustness was assessed using the Louvain community detection method (resolution 0.5). The Louvain detection was performed using unsupervised clustering, where the algorithm does not use labels. Instead, it groups data points based on intrinsic patterns or similarities in the data to capture cell-type heterogeneity.

**Graph Convolutional Neural Network (GCN):** A Graph Convolutional Network (GCN) was implemented with PyTorch Geometric to analyze single-cell gene expression data, as adapted from Wang et al. (2021)[b]. A k-Nearest Neighbor (k-NN) graph was constructed using sklearn to capture cell-cell similarity (10 neighbors per cell). The GCN had two graph convolution layers: the first mapped input features (number of genes) to 64 hidden units with ReLU activation, and the second predicted 3 clusters of a trilineage (i.e., two NSC-derived lineages: OPC-like, and NPC-like states, and a mesenchymal-like lineage). Training was performed over 100 epochs using the Adam optimizer with a learning rate of 0.01, and cross-entropy loss optimized cluster assignments. Post-GCN training, Louvain community detection (community package) was applied to the k-NN graph to refine clustering and identify predictive features of cluster transitions.

**Graph Attentional Network (GAT):** A Transformer-based Graph Attention Network (GAT) was implemented with PyTorch Geometric. A k-NN graph was constructed with 10 neighbors per cell. The model employed TransformerConv layers with 4 attention heads in each layer: the first layer mapped input features (number of genes) to 64 hidden dimensions, and the second layer predicted 3 clusters. Training was performed over 100 epochs using the Adam optimizer (learning rate = 0.01) and Cross Entropy Loss. Post-training, Louvain community detection was applied to the k-NN graph to identify transcriptional modules and transition genes steering phenotypic switches. The loss function had to be minimized to below 0.01. Using GCN and GAT feature importance markers, we performed Singular-Value Decomposition and Spectral clustering as dimensionality reduction to enhance structure detection in gene expression state-space, followed by network analysis.

**Spectral Clustering** identifies clusters by constructing a similarity graph and partitioning its Laplacian matrix to group nodes with high intra-cluster similarity. We first constructed a k-NN graph and applied SpectralClustering from sklearn (10 clusters), and extracted top genes based on mean expression within each cluster.

**Singular Value Decomposition (SVD):** performs dimensionality reduction by decomposing the expression matrix into principal components, revealing dominant patterns in expression space. Truncated SVD using the sklearn package was applied to each Louvain cluster, retaining 10 components, and the top 10 genes were identified from loading magnitudes for each component and ranked by statistical significance.

**Network Construction and Centralities**: Gene Regulatory Network (GRN) construction was based on cosine similarity (using sklearn) for the SVD and spectral clustering markers. Centralities (betweenness, degree, eigenvector, closeness) were calculated using networkx centrality functions. All feature importance genes across all algorithms obtained a $p<0.05$.

**BDM: Block Decomposition Method (BDM)** was used to quantify shifts in gene expression after perturbations using the pybdm library, which calculates complexity by partitioning the binary matrix into blocks (Zenil et al., 2019). Gene expression data was normalized using standardized and binarized with a threshold of 0.1, transforming values above the threshold to 1 and others to 0. For each dataset, DEGs were filtered using corresponding lists and their expression matrices. The BDM shift for each gene was calculated by perturbation analysis of its expression rows and comparing the perturbed matrix's BDM with the initial value. The results included the top 10 and bottom 10 genes with the highest BDM shifts. Since the direction (positive or negative) can both be useful for therapeutic interventions (e.g., inhibitor or upregulate for a shift in network complexity), both with the highest magnitude of BDM shift are reported. Gene set enrichment and functional validation of the identified markers were evaluated using g:Profiler and the GeneCards database.

# RESULTS

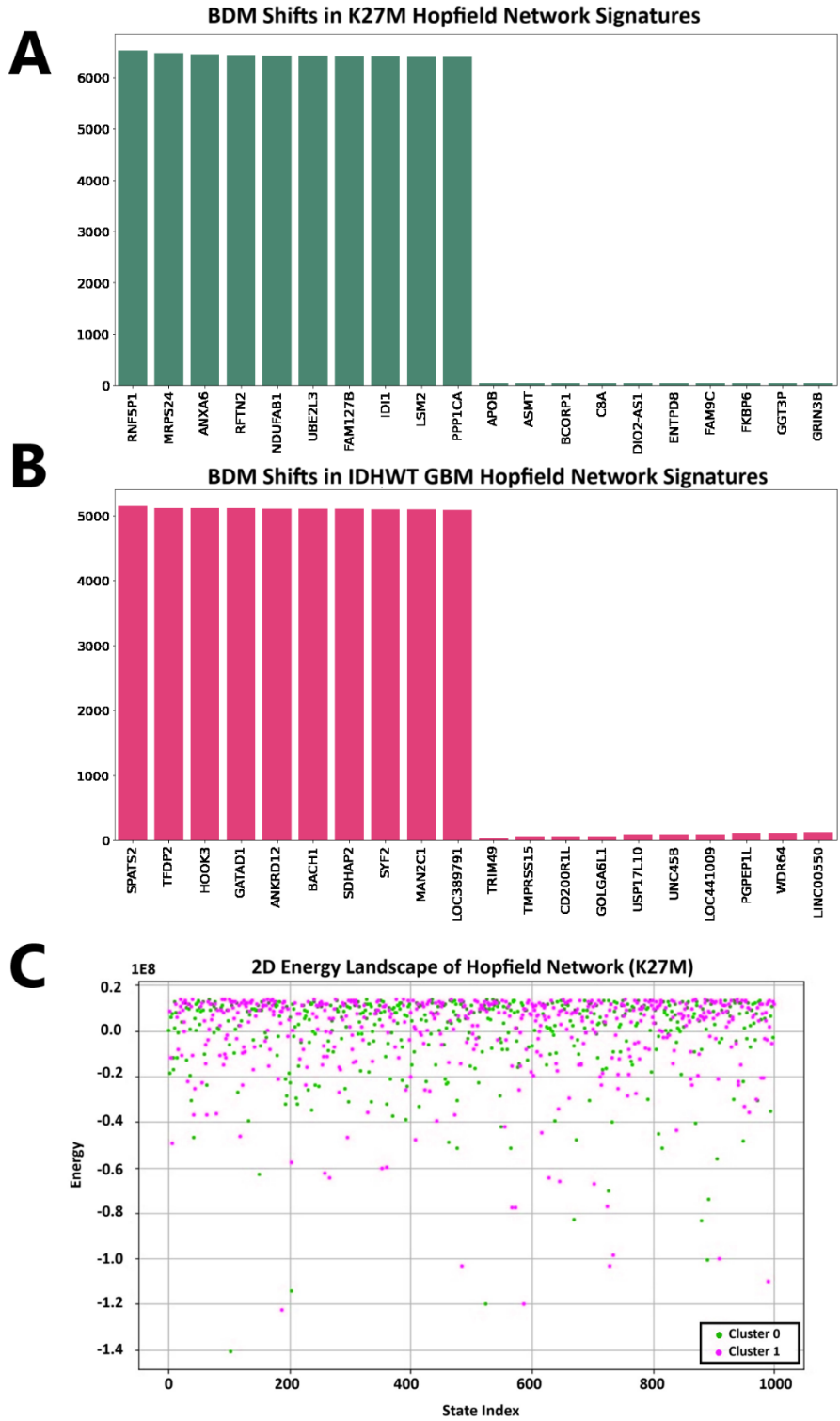

Figure 1. Hopfield Network Features. **A)** BDM shifts for key genes in K27M Hopfield network signatures highlight critical nodes driving phenotypic stability and state transitions. **B)** BDM shifts in IDHWT GBM

Hopfield network signatures. C) The 2D energy landscape of the K27M Hopfield network shows state attractors (clusters) representing stable phenotypic states and transitions in energy space. The model inferred the dynamics between two unsupervised clusters.

*Hopfield Energy Networks Reveal Metabolic Regulation, Chromatin Remodeling, and Developmental Dysregulation as Key Drivers of pHGG Cell Fate Transitions*

In K27M, the largest BDM shifts corresponded to ribosomal protein-encoding genes, with ANXA6 (involved in exocytosis and therapy resistance), NDUFAB1 (mitochondrial respirasome activity), and UBE2L3 (ubiquitination of p53 and NF-kB, linked to myelopoiesis) playing key roles (GeneCard Database) (Figure 1A). Most signatures highlighted metabolic functions and protein machinery regulation. PDGFRA and GAPDH were also observed in the top 30 BDM shifts. However, among the top feature importance genes, KDM4C, FKBP5, HES7, POU2F3, and MAPKAPK5 were found, and gene set enrichment analysis using g:Profiler identified Hippo signaling pathways (e.g., PAK1 and MOB1B). ZBTB33 (Kaiso), was also a common transcription factor (TF) in the top 30 feature importance genes in g:Profiler analysis. Notably, KDM4C is a histone demethylase affecting H3 lysine residues in H3K9 and H3K36 (Cascante et al., 2014), while FKBP6 and GRIN3B (NMDA-glutamate receptor) had the lowest BDM shifts.

In IDHWT, key genes with BDM shifts include SPATS2 (spermatogenesis regulation), TFDP2 (E2F transcription factor involved in cell cycle dynamics), HOOK3 (microtubule tethering and motor adapter proteins), and GATAD1 (Figure 1B). According to UniProt and GeneCards databases, GATAD1 is recruited to H3K4 methylated regions of the chromatin (Kana et al., 2020). Other notable genes include ANKRD12 (HDAC recruitment to nuclear receptors) and BACH1 (regulator of the NFE2FL2 oxidative stress pathway). Among salient features, SEZ6L, RAB41, and FXYD5 were identified with the highest feature importance scores, alongside chromatin remodeling genes such as BRD1, ING5, and histone markers (HIST1H2AD, HIST2H2AA3/4). CNTN2, involved in axon guidance, was also highlighted among the highest feature-importance genes. Interestingly, the top 10 lowest BDM shifts included sperm/testis antigens and lncRNAs, possibly indicating developmental dysregulation in these networks.

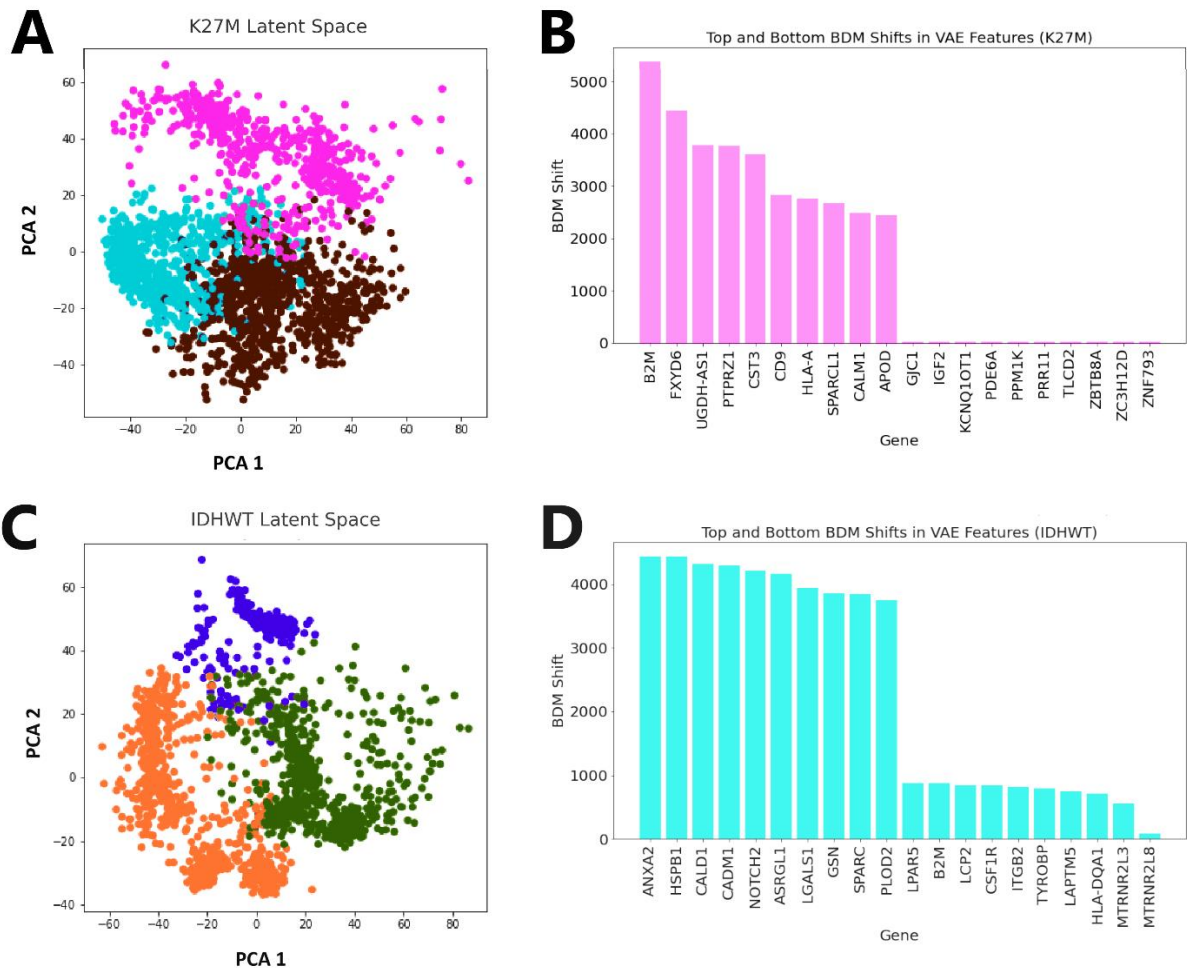

Figure 2. Variational Autoencoders (VAE).

*VAE Reveals Calcium Dynamics-Dependent Morphogenesis and Transition Genes*

In K27M, the highest BDM shift autoencoder features included FXYD6, B2M, PTPRZ1, and CD9, and genes with notable roles in calcium dynamics (e.g., CALM1, SPARCL1) and ECM remodeling. Other prominent signatures included PDGFRA, CST3, NRN1, GAPDH, METTL21A, IGF2, MEIS1/2, and SAPRCL1 (Figure 2B). In IDHWT glioblastoma, key BDM shift genes included NOTCH2, DKK3, ANXA2, LAPTM5, CX3C1, PTN, CD44, SOX11, and GFAP. SALL3, CADM1, and CADM2 were also observed which exhibit roles in cytokine regulation and natural killer cell targeting (Figure 2D). Calcium-dependent growth processes were seen with CALD1 and ANXA2, while ECM and cytoskeletal regulators like GFAP (astrocytic marker), DPP6, SPARC, and GSN were seen among the top feature importance markers drive morphogenesis. Shared signatures across gliomas included B2M, calmodulin, and markers of calcium dynamics.

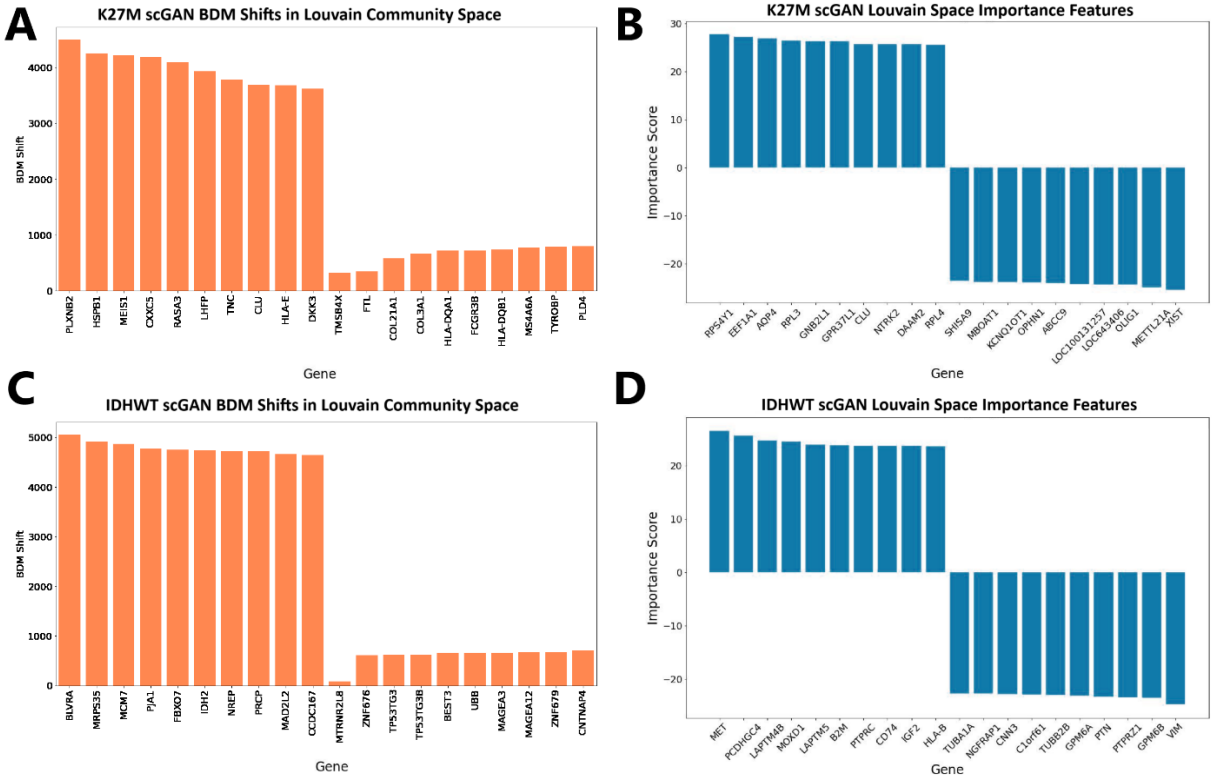

Figure 3. Generative Adversarial Networks (GAN): BDM Shift Signatures and Importance Features.

*GAN Highlights Metabolic Reprogramming and Immune Signatures in Glioblastoma Subtypes*

In K27M, the highest BDM shifts included PLXNB2, which according to the GeneCard Database regulates axon guidance and cancer invasion via RHO/RAC1/CDC42 and ERBB2 pathways, while CXXC5 and RASA3 contribute to immune signaling, myelopoiesis, and MAPK pathway activation (Figure 3A). Other key genes included TNC (ECM and synaptic plasticity), DKK3 (embryonic tissue differentiation via Wnt antagonism) (Kikuchi et al., 2022), and CLU (tumor immune infiltration) (Ren et al., 2023). Shared pathways among the two glioma subtypes involved cytoskeletal dynamics, ECM remodeling, and Wnt signaling. MEIS1 was also observed as a feature importance marker, which functions as a cofactor of HOXA7 and HOXA9, developmental genes involved in hematopoiesis, neurogenesis, and myeloid leukemias (Thorsteinsdottir et al., 2001). CXXC5 is also involved in immune signaling and a mediator of the BMP4-mediated canonical Wnt signaling activation in neural stem cells (NSCs). Positive importance genes included RPS4Y1 and EEF1A1 (protein synthesis, IFNG-linked immunity), AQP4 (astrocytic nutrient coupling at the blood-brain barrier), NTRK2 (BDNF receptor driving cell survival and synaptic plasticity), GNB2L1 (tumor growth and angiogenesis), GPR37L1 (ERK/MAPK regulation), and DAAM2 (Wnt signaling and actin dynamics). Negative importance features included OLIG1, METTL21A, XIST, and KCNQ1OT1, wherein OLIG1 is a critical pHGG marker (Figure 3B). Negative feature importance indicates that the feature negatively contributes to the model's predictions, suggesting the marker might be shared by more than one cluster among the Louvain communities. This suggests that features such as OLIG1 are critical hallmark signatures of pHGGs shared by collective cell dynamics (i.e., expressed in different phenotypic clusters).

In IDHWT, the highest BDM perturbations including mitochondrial bioenergetics and metabolism are highlighted by BLVRA (ROS and heme metabolism), MRPS35 (mitochondrial protein synthesis), and IDH2 (metabolic reprogramming) (Figure 3C). The presence of IDH2 is a metabolic driver gene validation of the IDHWT glioblastoma. MCM7 and PJA1 regulate DNA replication and ubiquitin-mediated chromosomal dynamics, while FBXO7 is involved in hematopoiesis. Key feature importance genes included MET, B2M, LAPTM5, MOXD1, and IGF2, revealing metabolic reprogramming and immune-related processes driving glioma progression. Positive importance genes included MET (EMT and morphogenesis via c-Met activation), PCDHGC4 (cell adhesion in the brain), IGF2 (tumor growth via mitogenic signaling), LAPTM4B, MOXD1, PTPRC, CD74, and B2M. Negative importance genes included TUBBA1A, PTPRZ1, GPM6A/B, and PTN, suggesting these features are co-expressed in distinct Louvain community clusters (Figure 3D). Collectively, these findings highlight that RHO, WNT, and metabolic dynamics shape both glioma systems.

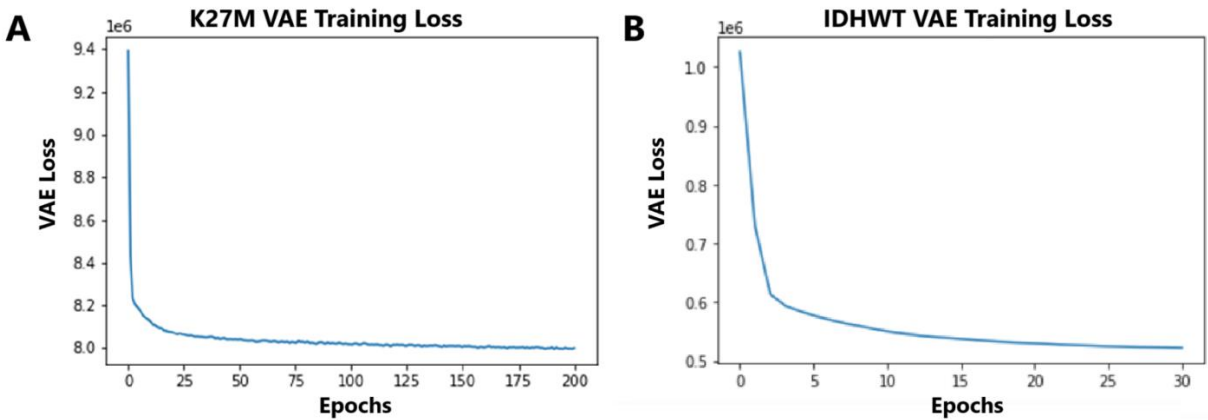

**Figure 4. Training Loss functions of the VAE Generator in GAN.** A) K27M stabilizing within 50 epochs of training. B) IDHWT shows a steep decline within the first 10 epochs, stabilizing by ~20 epochs indicating a faster convergence compared to K27M. The loss functions demonstrate that the generative model can well distinguish the three Louvain community clusters. Similar loss functions were observed for other algorithms such as the GCN and GAT.

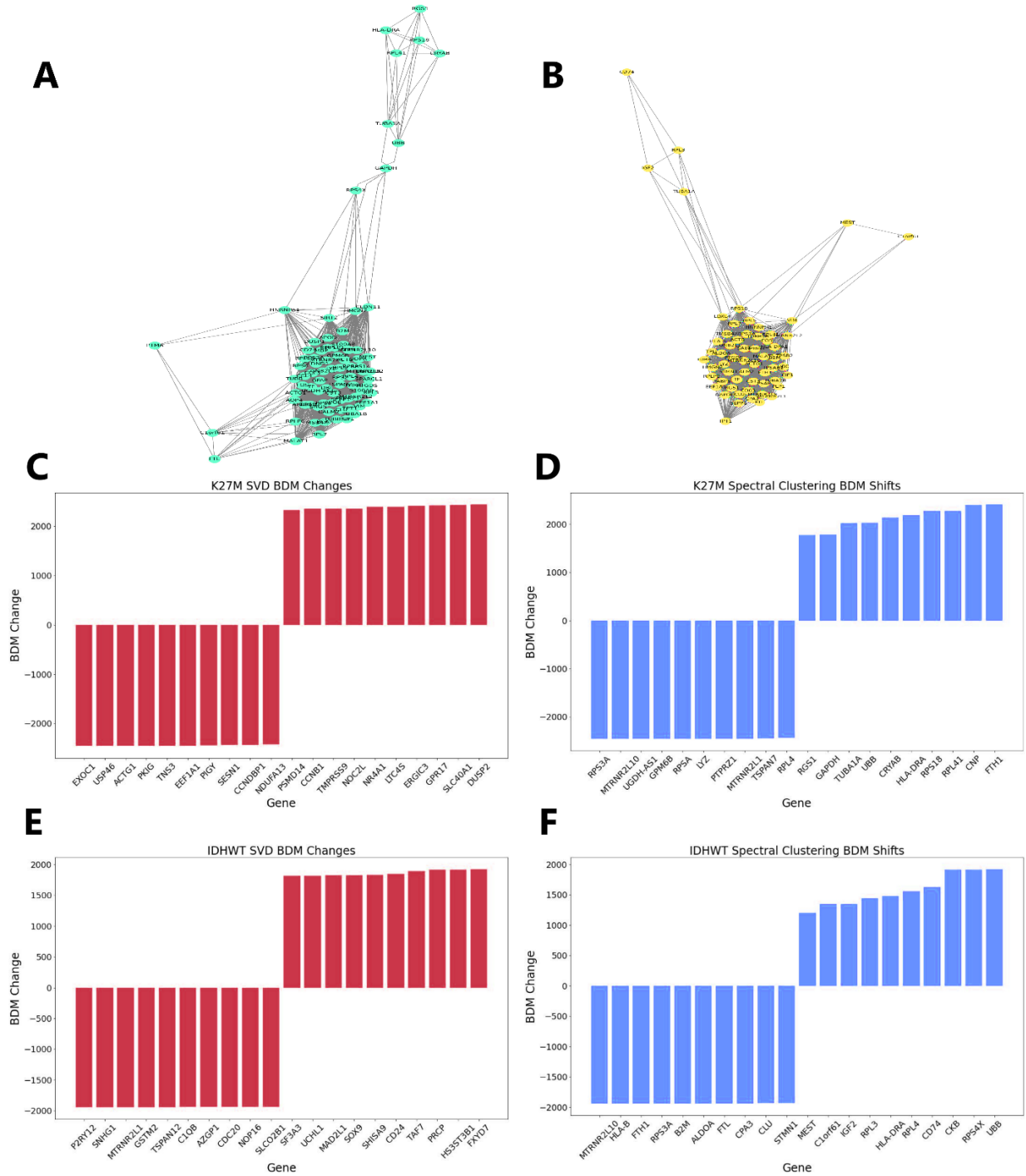

Figure 5. Graph Convolutional Networks (GCN). A) Network structure of Spectral cluster markers for K27M using GCN. B) Spectral clustering Network for IDHWT. C) K27M SVD BDM shifts. D) K27M Spectral clustering BDM shifts. E) IDHWT SVD BDM shifts. F) IDHWT Spectral clustering BDM shifts.

**GCN Reveals Extracellular Matrix, Metabolic, and Immune Signatures Driving Glioblastoma Progression**

**K27M SVD**: Top BDM shifts using SVD involved DUSP2, SLC40A1, and CCNB1, highlighting MAPK inactivation, iron homeostasis, and M-phase progression (Figure 5C). Within the top 20 BDM shifts, we also observed PDGFRA and FKBP2. DUSP2 inactivates MAP kinases via dephosphorylation. SLC40A1 is involved in cellular iron homeostasis. Cytoskeletal proteins (ACTG1) and mitochondrial components (NDUFA13) dominate the lowest BDM shifts, indicating a role in cytoskeletal dynamics and metabolic stress responses. **K27M Spectral Clustering:** The highest BDM shifts using spectral clustering included FTH, CNP, HLA-DRA, and FLRT (angiogenesis, epigenetic regulation, and protein synthesis). Positive features highlight roles in ECM remodeling (CLDN11) and hypoxia-driven processes (FLT), while negative shifts included ribosomal proteins (MTRNR2L1/10) and cytoskeletal markers (PTPRZ1). Only 20 genes had positive BDM shifts including HNRNPA1 and SIRT2 (Figure 5D). CLDN11 is involved in oligodendrocyte migration and blood-testis barrier regulation, while SIRT2 is a NAD+-dependent deacetylase with mixed implications in gliomas (McCornack et al., 2023). Most of the functional validations provided throughout are derived from g:Profiler and GeneCard Database.

**IDHWT SVD:** Top BDM shifts in SVD feature space included genes such as FXYD7, SOX9, SHISA9, HS3ST3B1, and MAD2L1, which indicate ion transport, differentiation, dynamics, and mitotic checkpoint regulation (Figure 5E). For instance, HS3ST3B1 catalyzes the transfer of sulfate to specific glucosamine residues in heparan sulfate. Negative shifts involved immune signaling (C1QB) and mitotic control (CDC20), alongside metabolic and neuronal differentiation regulators (ITM2C, FHL2). **IDHWT Spectral Clustering**: Top BDM shifts via spectral clustering included UBB, CKB, IGF2, and CD74, reflecting brain metabolism, MHC signaling, and oncogenic growth. Lowest shifts (MTRNR2L10, B2M, and ALDOA) underscore immune and glycolytic functions and ECM remodeling (CPA3, CLU). MTRNR2L1/2, EIF1, TUBB, TUBA1A, and TPT1 (Figure 5F). Most of these markers suggest neuronal lineage differentiation preference (Eze et al., 2021).

| Model | Network | Betweenness | Degree | Eigenvector | Closeness |
|---|---|---|---|---|---|
| IDHWT | Spectral | PMP2 | ITM2C | ITM2C | LINC00152 |
| IDHWT | Spectral | CMSS1 | NOP16 | NOP16 | ITM2C |
| IDHWT | Spectral | LINC00152 | ACTB | ACTB | NOP16 |
| IDHWT | SVD | PMP2 | ITM2C | ITM2C | ORC5 |
| IDHWT | SVD | CMSS1 | NOP16 | NOP16 | LINC00152 |
| IDHWT | SVD | LINC00152 | ACTB | ACTB | ITM2C |
| K27M | Spectral | GAPDH | FOS | RPS3A | SIRT2 |
| K27M | Spectral | SIRT2 | BCAN | MTRNR2L1 | RPS3A |
| K27M | Spectral | TUBA1A | AQP4 | ACTB | MTRNR2L1 |
| K27M | SVD | TRIM23 | EEF1A1 | EEF1A1 | EEF1A1 |
| K27M | SVD | SH3BGRL | EXOC1 | USP46 | USP46 |
| K27M | SVD | H3F3A | USP46 | EXOC1 | EXOC1 |

**Table 1. Centrality Measures for scGCN Markers by Network Type (SVD or Spectral Clustering).** In IDHWT, the highest centralities included PMP2, LINC00152, ACTB, and IT2MC, K27M highlighted MTRN2L1/2, SIRT, ACTB, H3F3A, and SH3BGRL, among the top centralities. These findings suggest

developmental features and morphogenetic patterns across both pHGG subsystems. For instance, LINC00152 is linked to EMT transitions, while ITM2C is involved in the negative regulation of neuronal differentiation and neuron projection development with TNF signaling. H3F3A is the (epigenetic) driver mutation in K27M gliomas.

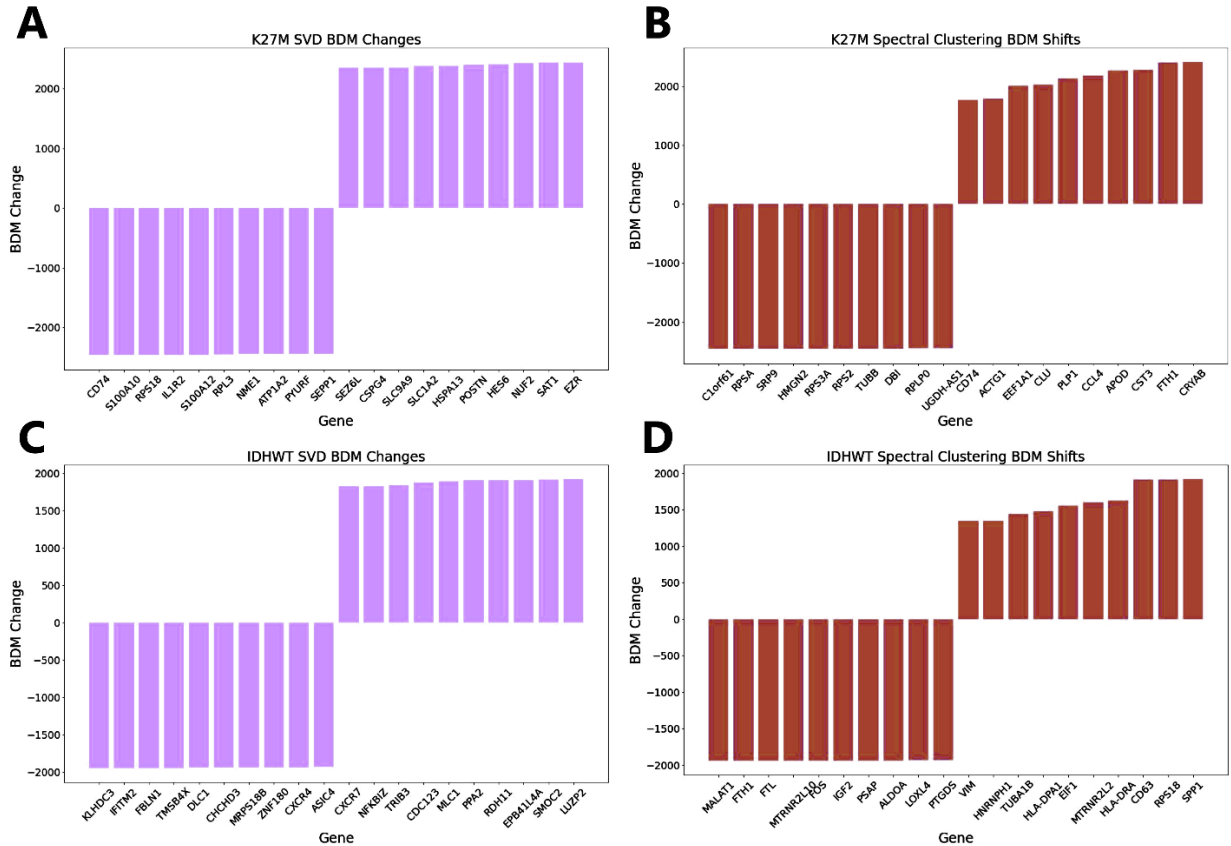

**Figure 6. Graph Attentional Networks (GAT) BDM Perturbation Analysis**. A) SVD-based BDM shifts in K27M. B) Spectral Clustering BDM shifts in K27M. C) SVD-based BDM shifts in IDHWT glioblastoma. D) Spectral clustering-based BDM shifts in IDHWT.

*BDM Analysis on GAT Importance Features Reveals Shared Patterns of Cytoskeletal, Immune, and Metabolic Regulation in Glioblastoma Subtypes*

**K27M SVD:** Highest positive BDM shifts for single gene signatures included EZR, SAT1, NUF2, HES6, POSTN, HSPA13, SLC1A2, SEZ6L, CSPG4, SLC9A9 (Figure 6A). HES6 encodes a transcriptional repressor involved in neurogenesis, glioma differentiation, and somitogenesis (Couturier et al., 2020). SLC1A2 encodes a glutamate transporter that clears extracellular glutamate from synapses and limits glutamatergic neurotransmission in the brain. SEZ6L encodes a membrane protein involved in developmental processes but its exact function is unclear. CSPG4 encodes a type 1 transmembrane proteoglycan involved in cell motility, proliferation and adhesion in various cell types such as glia and melanoma. EGR2, FXYD6, S100B, XIST, and S100A9 were also observed in top BDM shifts in both SVD and Spectral clustering for K27M.

**K27M Spectral:** Highest BDM shifts were CRYAB, FTH1, CST3, APOD, CCL4, PLP1, CLU, EEF1A1, ACTG1, CD74 (Figure 6B). CCL4 encodes a small cytokine that belongs to the CC chemokine family and functions as a chemoattractant for immune cells during inflammation. PLP1 encodes proteolipid protein 1, which is the most abundant protein in the myelin sheath of neurons (Table 1). APOD encodes apolipoprotein D, which is involved in lipid transport, and modulation of immune-inflammatory responses. Again we see FTH1 and EEF1A1. FTL was also observed in the top 20 BDM shifts.

**IDHWT SVD:** The highest BDM shifts included LUZP2, SMOC2, EPB41L4A, RDH11, PPA2, MLC1, CDC123, TRIB3, NFKBIZ, and CXCR7 (Figure 6C). Notably, some other markers such as NUTF2, TNR, DCTN3, and DBI were also among the top 20, revealing neuro-immune processes and metabolic regulatory roles.

**IDHWT Spectral:** Highest BDM shifts were SPP1, RPS18, CD63, HLA-DRA, MTRNR2L2, EIF1, HLA-DPA1, TUBA1B, HNRNPH1, VIM (Figure 6D). SPP1 regulates cytoskeletal-matrix dynamics and is a cytokine that regulates interferon-gamma and interleukin-12.

| Model | Network | Betweenness | Degree | Eigenvector | Closeness |
|---|---|---|---|---|---|
| IDHWT | Spectral | TUBB | TUBB | RPL41 | TUBB |
| IDHWT | Spectral | C1orf61 | RPL41 | MTRNR2L1 | RPL41 |
| IDHWT | Spectral | ACTG1 | C1orf61 | ACTB | C1orf61 |
| IDHWT | SVD | NRP1 | HSPA13 | MTRNR2L8 | MTRNR2L8 |
| IDHWT | SVD | SCOC | ZNF180 | ZNF180 | ZNF180 |
| IDHWT | SVD | TUBB2B | MTRNR2L8 | CHIC2 | HSPA13 |
| K27M | Spectral | ACTG1 | FOS | RPSA | UBB |
| K27M | Spectral | UBB | GPM6B | SRP9 | RPSA |
| K27M | Spectral | EEF1A1 | DBI | RPS3A | SRP9 |
| K27M | SVD | ATP9B | RPL3 | RPL3 | RPL3 |
| K27M | SVD | DUSP1 | TXNL1 | S100A12 | TXNL1 |
| K27M | SVD | OLIG1 | S100A12 | IL1R2 | S100A12 |

Table 2. Network Centrality Measures for GAT Across pHGG Subsystems. Key centrality features across IDHWT and K27M highlight cytoskeletal dynamics (TUBB, ACTG1), ribosomal proteins (RPL41, RPL3), and mitochondrial genes (MTRNR2L1/8) as central in IDHWT, while K27M emphasizes ECM remodeling (UBB, GPM6B) and immune-inflammatory signaling (S100A12, IL1R2). Shared patterns include structural and metabolic regulation driving glioma progression. PRC1, METTL7B, and TOP2A were observed among the highest betweenness also for IDHWT spectral clustering. ADAMTS1 and CHCHD3 were also observed repeatedly among the centralities.

## DISCUSSION

### *Deep Learning-based Biomarker Discovery Unveil Immune, Metabolic, and Microenvironmental Remodeling in Maladaptive Tumor Evolution*

The artificial neural networks reconstructed the transition genes underlying the cell fate trajectories (state-attractors) from single-cell transcriptomics. The convergence of similar plasticity (transition) markers validates some ground truth in the predictive modeling. GNNs model cellular interactions as graph networks, while VAEs and GANs generate latent representations of cell states, and Hopfield networks identify basins of attraction (cell states) on an energy (epigenetic) landscape, offering insights into cellular decision-making (Wang et al., 2008; Li and Wang; 2013; Wu and Wang; 2014). BDM perturbation analysis and network centralities complement these algorithms by identifying critical genes that destabilize network dynamics when perturbed, highlighting key therapeutic targets for precision medicine. Together, these methods converged on shared neuro-immune-metabolic processes and extracellular matrix (ECM) or cytoskeletal remodeling features, which are central to tumor microenvironment evolution and phenotypic plasticity.

Our findings align with the results from Petralia et al. (2020) which show that BRAF mutations in the MAPK/MEK/ERK pathway are most frequent in low-grade gliomas, while TP53 and H3F3A mutations dominate high-grade DIPG gliomas. Similarly, PTPN11 mutations linked to the MAPK/RAS pathway have been reported in high-grade gliomas, as was observed herein (Petralia et al., 2020). In our present study, key differentiation pathways coordinating cell fate transitions indicate NSC (Radial glia) markers across both subsystems, and morphogen networks (developmental patterning genes) including WNT, MAPK/ERK, TNF, IGF2, and RHO/RAC signaling, along with epigenetic and immune-inflammatory signals. These network processes reveal the top-down control of tumor microenvironment dynamics in phenotypic plasticity. Further, these insights emphasize the role of immune evasion, metabolic reprogramming, and ECM remodeling in steering phenotypic transitions during tumor progression/invasion and could be repurposed towards identifying potential targets for precision immunotherapy or epigenetic reprogramming of cancers towards stable attractor states (i.e., differentiated, cell fate commitments).

### *Radial Glia Markers and Signaling Pathways Highlight Forebrain and PFC Neuronal Lineage Differentiation*

Many of the transition genes identified in this study were recently found to be markers of adult radial glia (NSCs) of the subventricular zone (Baig et al., 2024). These include SPARCL1, FABP7, GFAP, BCAN, HES6, PTN, CLU, PTPRZ1, MEG, C1orf61, GFAP, SPARCL1, VIM, etc. (Baig et al., 2024). Most of the top gene centralities identified by spectral clustering using the GAN, GCN, and GAT networks align with these neurogenic radial glia (NSCs) and neuronal differentiation markers. For instance, MOXD1, PTPRZ1, FABP7, C1orf61, SOX11, and HES6 identified as transition genes herein were found to be key markers enriched in glioblastoma stem-like states, and in neurogenic radial glia of the central telencephalon (Pine et al., 2023; Baig et al., 2024) (Table 2). Thus, these cell state transition markers confer the plasticity and dedifferentiation in pHGG cellular (developmental) hierarchies. Our findings suggest that pHGG cell fate choices favor higher-order, neocortical differentiation. Markers such as HES5/6, SOX9, and MEIS1 observed in our study are also signatures of the prefrontal cortex (PFC) development from radial glia/NSCs (Dulken et al., 2017; Eze et al., 2021). The PFC is the most recent evolutionary development of the neocortex responsible for higher-order cognitive functions such as decision-making, problem-solving, and teleonomic behaviors. Further, markers such as H2AFZ, CALM1, and LAPTM4 seen in our findings were the top enriched markers in the telencephalon (i.e., the anterior part of the forebrain) from radial glia differentiation (Eze et al., 2021). Dynamic oscillations in Wnt, Fgf, Hes5/6, and Notch signaling pathways drive neocortical development from radial glia (Eze et al., 2021;

Baig et al., 2024), as coordinated by the clock and wavefront model (Gibb et al., 2010; Sonnen et al., 2018). These oscillations are also critical for EMT transitions (plasticity dynamics). Notch signaling preserves stem cell pools and limits premature neurogenesis, while Wnt modulates medial-lateral patterning and progenitor pool maintenance, collectively coordinating early brain patterning and differentiation (Eze et al., 2021). Critical markers of amyloid fiber formation and *basal ganglia* (i.e., the 'reptilian' brain) *dysgenesis* were also observed in support of this (Bennett, 2023), with markers including CST3, B2M, LYZ, UBB, and UBC. B2M, for instance, is a prognostic serum marker for gliomas (Liu et al., 2024). Thus, collectively our findings suggest pHGGs express radial glia markers favoring forebrain/PFC neuronal development.

**Major Transition Patterns and Lineage Markers Observed in IDHWT Gliomas**

In IDHWT gliomas, significant signatures include histone modification, chromatin remodeling, and immune response markers. Genes like SPATS2 and TFDP2 highlight roles in cell cycle dynamics and oxidative stress regulation (e.g., BACH1). Metabolic reprogramming is evident through GFAP, DPP6, and LRRN1 genes, which regulate potassium channels, ECM remodeling, and synaptic signaling. Interestingly, calcium dynamics (e.g., CALD1, ANXA2) are a recurring feature, alongside cytoskeletal markers like MAP1B and TAGLN3. These patterns suggest adaptive neuro-immune reprogramming and transitions in cancer morphogenesis. The highest BDM shifts included signatures such as DKK3, and GFAP (Figures 2 and 3). In contrast, the lowest shifts corresponded to cytoskeletal and ECM components such as TUBBA1A, contributing to tumor network resilience and stability.

**Major Differentiation Patterns and Lineage Markers Observed in K27M Gliomas**

K27M gliomas are characterized by ribosomal proteins and metabolic machinery (e.g., FXYD6, NDUFAB1), highlighting mitochondrial activity and translational regulation. Key markers such as DKK3, ANXA6, and UBE2L3 (ubiquitination and NF-κB regulation) underline roles in protein transport and signaling pathways. H3F3A, the histone mutation marking K27M, and SIRT2, a histone deacetylase (HDAC), emerged among the top network centralities in K27M using the GCN algorithm, demonstrating its epigenetic subtype (functional) validation (Table 1). The Hippo signaling pathway was commonly implicated, with top genes like KDM4C and CAMK2B, involved in histone modification and calcium signaling. Regulatory hubs such as Kaiso (ZBTB33) might interconnect Wnt and HDAC pathways. The highest BDM shifts included calcium-dependent proteins and ECM markers (e.g., SPARCL1, CALM1), while the lowest shifts (e.g., GRIN3B) (Figure 1) suggested endogenous bioelectric network components. These patterns emphasize phenotypic plasticity and network destabilization driving glioma progression.

Furthermore, the GCN and GAT spectral clustering gene centralities, predominantly reflecting radial glia and neuronal markers (Baig et al., 2024), reveal key overlaps in gene expression. For K27M, the top shared genes include SIRT2, RPLP0, CALM2, DBI, AQP4, CLDND1, RPL3, MTRNR2L10, RPL4, and UGDH-AS1, while for IDHWT, they include GAPDH, PTGDS, MTRNR2L1, EEF1A1, ACTB, and CST3 (Figures 5 and 6). Across both groups, the common genes ACTB, PTGDS, and MTRNR2L1 emerge as central markers.

*Markers of Iron Metabolism, Extracellular Vesicles, and Neurodegenerative Network Dynamics in scGCN IDHWT Glioblastoma*

In scGCN analysis of IDHWT glioblastoma, resilience markers from SVD include MTRNR2L1 and P2RY12, while FTH1, FTL, and ALDOA dominate spectral clustering. Positive BDM shifts were observed in SOX9, FXYD7, and IGF2, while UBB was identified across both glioma systems (Table 2). Iron metabolism signatures, including FTL, FTH1, and ferroptosis pathways, highlight the role of ferric ion binding in tumor progression (Zhang et al., 2022). g:Pofiler-based Gene set enrichment analysis revealed extracellular vesicle/exosome pathways (P=2.52E-22), with genes such as STMN1, EEF1A1, ACTB, and TPI1 driving this process. Distinct negative markers, including ACTG1 (actin dynamics), ALDOA (glycolysis), CD99, and CD63 (tumor progression and immunotherapy target), were also observed, alongside CD74 and MEST (linked to mesodermal dysregulation). Further, FTL, FTH1, GAPDH, and ALDOA highlight iron metabolism as a therapeutic vulnerability. GFAP identified as a feature by VAE is also a blood-based Alzheimer's biomarker (Sintini et al., 2024).

*Convergent Pathways in IDHWT and K27M: Iron Metabolism, Mitochondrial Regulation, and Neurodevelopmental Transitions in GCN and GAT*

In the GCN analysis, we identified FTH1, SLC40A1, CRYAB, and UBB as markers of iron metabolism in K27M (Figure 5; Table 2). Gene set enrichment analysis for GAT in both K27M and IDHWT revealed ferric and ferrous iron binding (TF, FTL, FTH1), indicating a shared metabolic signature across glioma subtypes. In K27M, the GAT algorithm highlighted SEZ6L, SLC1A2, SLC9A9, and HES6 as key neurodevelopmental and bioelectric markers (Figure 6). These genes are involved in synaptic development (SEZ6L), glutamate clearance (SLC1A2), pH regulation (SLC9A9), and neural differentiation (HES6), pointing to neuronal differentiation pathways. Additionally, GCN analysis identified CNP, PLP1, SIRT2, and GPM6B as neurodevelopmental markers in K27M.

*MAPK/ERK and Wnt Signaling Feedback Loops as Drivers of Differentiation Hierarchies and Grade-dependent Glioma Plasticity*

MAPK/ERK pathways and WNT signaling, critical for proliferation-plasticity feedback dynamics in cancer stem cells, were observed among the identified transition markers (Gibb et al., 2010; Sonnen et al., 2018; Roberts et al., 2023). Wnt signaling was identified by markers such as DKK3, CXXC5 (Figure 3), and DAAM2. Key MAPK pathway markers included DUSP1/2, a negative regulator of MAPK/ERK, RASA3, CXXC5 (i.e., mediates BMP4-induced canonical Wnt and MAPK signaling in NSCs), HES6, a transcriptional repressor involved in neurogenesis, and GRIN3B, connecting calcium dynamics and MAPK signaling during neuronal differentiation. We predict this feedback loop might underlie the differentiation hierarchies between low-grade and high-grade gliomas. In speculation, the epigenetic or histone code underlying H3K4, H3K9, and H3K36 methylation dynamics may further steer these hierarchical patterns.

*Cancers as Cell Identity Disorders: Behavioral Dynamics and Maladaptive States*

Our findings suggest that pHGG ecosystems are 'cellular identity disorders', or morphogenetic (patterning) disorders, characterized by disrupted developmental hierarchies. These tumors navigate *critical transitions* in cell fate choices, switching phenotypes as an evolvability engine to counteract environmental stressors and uncertainty (Shin and Cho, 2023). The excessive plasticity, or metastability, causes tumor cells to remain trapped in unstable critical states, resulting in *maladaptive behaviors*. Our findings support the theory that NSCs in the subventricular zone (SVZ), identified as cells of origin for glioblastomas (Ocasio et al., 2023; Wang et al., 2024), drive tumor development dysregulated developmental programs at the bifurcation point of patterning processes such as somitogenesis, neural crest and neural tube formation (Uthamacumaran, 2024). While K27M gliomas are currently held to be derived from pre-OPC or OPC lineages as their cell of origin (Andrade et al., 2020), our findings suggest that both pHGG subtypes seem to favor neuronal identities by their upregulation of neuronal lineage markers and neuronal differentiation programs.

*Reprogramming Cancer Plasticity: A Suggested Interpretation*

pHGG behavior reflects a continuum of adaptive yet pathological states, overlapping with precision biomarkers of neurodevelopmental or neuropsychiatric disorders like schizophrenia, autism, and bipolar disorder. In specific, neuropsychiatric susceptibility genes such as FXYD6, SEZ6L, SOX11, etc. identified in our previous study (Uthamacumaran, 2024), were re-validated here in deep learning-based biomarker discovery, further supporting the neurodevelopmental and neuro-immune-glioma axis Zhong et al., 2011; Xu et al., 2013; Najjar et al., 2013). For instance, FXYD6 encodes an ion transporter debated as a susceptibility gene for schizophrenia, highlighting the parallels between glioma plasticity and maladaptive behaviors in neuroinflammatory environments. FKBP5 observed in our study as a transition marker is also associated with aberrant glucocorticoid receptor signaling and disrupted hypothalamus-pituitary-adrenal (HPA) axis in traumatic disorders (Roy et al., 2010). Further, the emerging role of calcium dynamics in neuroplasticity, neurotransmission, and antidepressants such as vortioxetine in modulating glioma NSC niches underscores the shared neuropsychiatric and cancer axis (Lee et al., 2024). In proposition, reprogramming glioma phenotypes toward neuronal-like lineage differentiation (i.e., stable cell fates) can repurpose their plasticity towards cancer reversal to benign-like states, offering potential therapeutic avenues. In evidence, Liao et al. (2024) recently demonstrated that the exogenous application of NRXN1 could differentiate adult glioblastoma towards neuronal-like lineage differentiation. This finding highlights that bioelectric networks can reprogram the epigenetic plasticity of cancer cell fates, and more interestingly, as addressed by the authors, NRXN1 is a functional biosignature of schizophrenia and autism spectrum involved in glioblastoma cell fate modulation (Liao et al., 2024). Therefore, transition therapies targeting the immune-inflammatory microenvironment and developmental blockages may aid in overcoming gliomas' arrested differentiation and maladaptive behaviors, steering them toward functional neuronal fates ((Solé and Aguadé-Gorgorió , 2021; Aguadé-Gorgorió et al., 2022; Shin and Cho, 2023).

**Limitations and Prospective Studies**

This study represents the first large-scale Generative AI and graph-based deep-learning analysis of single-cell transcriptomics in pediatric high-grade gliomas, as pattern discovery and causal inference algorithms (BDM perturbation analysis) to uncover explainable features (i.e., clinically actionable targets such as gene network signatures and biomarkers) as therapeutic vulnerabilities (Lusch et al., 2018; Luo et al., 2020). While interpretable deep learning techniques like feature importance offer valuable insights, some data-driven limitations remain. Batch effects were minimized by quality control and data pre-processing across multiple patients for each pHGG subtype. However, the lack of temporally resolved (longitudinal) gene expression limits our analyses to pseudotemporal inference. With time-series gene expression, derived from xenograft models or organoids, we could attempt optimal transport theory, differential equations-based neural nets, or recurrent neural networks for attractor inference (Schiebinger et al., 2019). Algorithms such as Long Short-Term Memory (LSTM) networks, Reservoir Computing, and Transformers could address in future studies. Further, deep learning models often operate as "black boxes", and their generalizability is questionable to larger patient cohorts. However, the validation of similar markers from previous experimental studies and across distinct algorithms addresses this concern. This was further demonstrated by the minimized loss function (MSE) (Figure 4). Prospective studies should focus on enhancing model interpretability to link feature discovery directly to actionable biological targets. Additionally, expanding the use of graph neural networks (GNNs) could improve the modeling of dynamic gene expression patterns within graph-structured systems.

**CONCLUSION**

pHGGs exemplify the complex interplay of neurodevelopmental arrest, immune-inflammatory adaptation, and plasticity-driven maladaptive behaviors (stress-response patterns). Deep learning-based biomarker discovery highlighted critical pathways and actionable therapeutic vulnerabilities, suggesting that these tumors, though trapped in unstable attractor states on their epigenetic landscapes, strive for neuronal differentiation amidst environmental stressors. This provides a compelling rationale for precision strategies targeting phenotypic plasticity's reprogramming potential.

Our findings suggest that targeting the predicted WNT-MAPK/ERK pathway feedback loop offers a route for *transdifferentiation therapy*, i.e., steering glioblastoma cells toward their favored neuronal lineage differentiation. Transdisciplinary *systems medicine* approaches, integrating data science with cancer multiomics, epigenetic reprogramming, and drug repurposing, could enable the reintegration of these pathological attractors (tumor ecosystems) into the physiological whole. Using these approaches, we identified immune, metabolic, epigenetic, and bioelectric signatures driving the collective intelligence of tumors in morphospace (i.e., 3D-patterning space). We also predicted epigenetic plasticity pathways underlying the plasticity transitions, such as GATAD1 (Figure 1) and KDM4C suggest roles for H3K4, H3K9, and H3K36 methylation, while H3F3A is associated with the K27M mutation (McCornack et al., 2023). Some studies reveal that NSC differentiation is regulated by KDM4C, which promotes neuronal differentiation through H3K9 demethylation while repressing astrocytic differentiation by demethylating H3K36 and inhibiting GFAP transcription elongation (Cascante et al., 2014). Our whole-systems or whole-person perspective of pHGG cybernetics resonates with the 'psychotherapeutic' paradigm seen in trauma-informed care and the biopsychosocial model of holistic healing practices. As such, our approach addresses pHGGs as cellular identity disorders to tackle the neurodevelopmental and neuropsychiatric dimensions of their maladaptive behaviors (Solé and Aguadé-Gorgorió, 2021; Aguadé-Gorgorió et al., 2022; Baig and Winkler, 2024).

In conclusion, systems medicine-driven AI approaches, such as GNNs and Generative AI have identified causal patterns and predictive biomarkers underpinning cell fate plasticity in pediatric glioblastoma and K27M DIPG. Our collective findings suggest that pHGGs are stuck in an OPC/NPC hybrid state while expressing NSC (radial glia) markers striving toward neuronal lineage differentiation (Wang et al., 2024; Baig et al., 2024). Further, the neurogenic radial glia markers expressed by both pHGG subtypes suggest neocortical neurons of the telencephalon and forebrain as their desired cellular identity. These findings pave the way for precision medicine strategies, including cellular reprogramming (cancer reversal) via transition or differentiation therapies, offering novel, effective, and potentially safer avenues for improving quality patient care in precision oncology.

**DATA AND CODE AVAILABILITY:**
https://github.com/Abicumaran/Glioblastoma_IV_DeepLearning

**DECLARATIONS:**

There are no disclosures or competing interests.